\documentstyle[12pt,a4wide]{article}

\begin{document}

\author{B. Linet \thanks{E-mail : linet@celfi.phys.univ-tours.fr} \\
\small Laboratoire de Math\'ematiques et Physique Th\'eorique \\
\small CNRS/UPRES-A 6083, Universit\'e Fran\c{c}ois Rabelais \\
\small Facult\'e des Sciences et Techniques \\
\small Parc de Grandmont 37200 TOURS, France}

\title{\bf Spin-weighted Green's functions \\
in a conical space}

\date{}
\maketitle

\begin{abstract}
We give an analysis of the spin-weighted Green's functions 
well-defined in a conical space.
We apply these results in the case of a straight cosmic string and in the Rindler
space in order to determine generally the Euclidean Green's functions for the
massless spin-$\frac{1}{2}$ field and for the electromagnetic field. We give also the
corresponding Green's functions at zero temperature. However, 
except for the scalar field, it seems that these Euclidean Green's functions 
do not correspond to the thermal Feynman Green's functions.
\end{abstract}

\thispagestyle{empty}
\newpage

A conical space is an Euclidean space with a conical-type line singularity which can be
described by the metric
\begin{equation}
\label{1}
ds^{2}=d\rho^{2}+B\rho^{2}d\varphi^{2}+(dx^{1})^{2}+\cdots +(dx^{n-2})^{2}
\end{equation}
in a coordinate system $(\rho ,\varphi ,x^{i})$,
$i=1, \dots ,n-2$, such that $\rho > 0$ and $0\leq \varphi <2\pi$, the hypersurfaces
$\varphi =0$ and $\varphi =2\pi$ being identified. 
Metric (\ref{1}) is characterized by a constant
$B$ supposed strictly positive; it reduces to the Euclidean metric for $B=1$.

In the massless case, the spin-weighted Green's functions ${\cal G}_{(s)}$, 
$s$ being integer or half-integer, obeys the equation \cite{aliev}
\begin{eqnarray}
\label{2}
\nonumber & & \left( \frac{\partial^{2}}{\partial \rho^{2}}+\frac{1}{\rho}
\frac{\partial}{\partial \rho}+\frac{1}{B^{2}\rho^{2}}\frac{\partial^{2}}
{\partial \varphi^{2}}+\frac{2is}{B\rho^{2}}\frac{\partial}{\partial \varphi}
-\frac{s^{2}}{\rho^{2}}+\triangle_{n-2} \right) {\cal G}_{(s)}= \\
& &-\frac{1}{B\rho}\delta (\rho -\rho_{0}) \delta (\varphi -\varphi_{0})
\delta^{(n-2)}(x^{i}-x_{0}^{i})
\end{eqnarray}
where $\triangle_{n-2}$ is the usual Laplacian operator. For $s$ half-integer, we 
suppose that ${\cal G}_{(s)}$ is antiperiodic in $\varphi$ with period $2\pi$ and
for $s$ integer that ${\cal G}_{(s)}$ is periodic in $\varphi$ with period $2\pi$. 
We impose that ${\cal G}_{(s)}$ vanishes when the points $(\rho ,\varphi ,x^{i})$ and
$(\rho_{0},\varphi_{0},x_{0}^{i})$ are infinitely separated. 
Furthermore, we require that ${\cal G}_{(s)}$ is bounded in the limit where
$\rho \rightarrow 0$. These conditions should be uniquely determined the solution.
We remark that ${\cal G}_{(s)}=\overline{{\cal G}}_{(-s)}$.

The spin-weighted Green's functions can be directly related to the Green's function
$D_{\Phi}$ of the Laplacian operator for metric (\ref{1}), {\em i.e.} 
equation (\ref{2}) with $s=0$, having 
the following conditions of periodicity in the coordinate $\varphi$
\begin{eqnarray}
\label{7}
\nonumber D_{\Phi}(\rho ,\varphi +2\pi ,x^{i})&=&\exp (2i\pi \Phi )
D_{\Phi}(\rho ,\varphi ,x^{i}) \\
\frac{\partial}{\partial \varphi}D_{\Phi}(\rho ,\varphi +2\pi ,x^{i})&=&
\exp (2i\pi \Phi )\frac{\partial}{\partial \varphi}D_{\Phi}(\rho ,\varphi ,x^{i})
\end{eqnarray}
and vanishing when the points $(\rho ,\varphi ,x^{i})$ and
$(\rho_{0},\varphi_{0},x_{0}^{i})$ are infinitely separated. We remark that 
$D_{\Phi}=\overline{D}_{-\Phi}$.
From equation (\ref{2}), it is easy to show that
\begin{eqnarray}
\label{9}
\nonumber & & {\cal G}_{(s)}=\exp [-iBs(\varphi -\varphi_{0})] \,
D_{1/2+Bs}(\rho ,\varphi -\varphi_{0},x^{i}) \quad s=\pm 1/2, \pm 3/2, \dots \\
& & {\cal G}_{(s)}=\exp [-iBs(\varphi -\varphi_{0})] \,
D_{Bs}(\rho ,\varphi -\varphi_{0},x^{i}) \quad s=0, \pm 1, \dots 
\end{eqnarray}
We point out that $D_{\Phi}$ depends only on the fractional part of $\Phi$, 
denoted $\gamma$ such that $0\leq \gamma <1$. Henceforth, we write $\gamma =[\Phi ]$.
A detailed analysis of the Green's function $D_{\Phi}$ has been done by 
Guimar\~aes and Linet \cite{gui}. In consequence, the study of the spin-weighted Green's
functions ${\cal G}_{(s)}$ results immediately from relations (\ref{9}).

Equation (\ref{2}) appears in the study of the field theories on the spacetime of a
straight cosmic string for different values of the spin. The Euclidean Green's
function $S_{E}$ for a massless spin-$\frac{1}{2}$ field is given by
\begin{equation}
\label{4}
S_{E}=\left( e_{\underline{a}}^{\mu}\gamma^{\underline{a}}\partial_{\mu}
+\frac{\gamma^{\underline{1}}}{2\rho} \right) 
\left( I\Re {\cal G}_{(-1/2)}+\gamma^{\underline{1}}\gamma^{\underline{2}}
\Im {\cal G}_{(-1/2)} \right)
\end{equation}
in polar vierbein $e_{\underline{a}}^{\mu}$ 
\cite{fro,lin}. The Euclidean Green's function $G_{E\mu \mu_{0}}$
for the electromagnetic field can be found by the following relation \cite{all}
\begin{equation}
\label{5}
G_{E\rho \rho_{0}}-\frac{i}{B\rho}G_{E\varphi \rho_{0}}={\cal G}_{(-1)}
\end{equation}
the other components being
\begin{equation}
\label{6}
G_{E\rho \varphi_{0}}=-\frac{\rho_{0}}{\rho}G_{E\varphi \rho_{0}} \quad {\rm and}
\quad G_{E\varphi \varphi_{0}}=B^{2}\rho \rho_{0}G_{E\rho \rho_{0}}
\end{equation}
Similar formulas with $s=2$ exist for the linearized gravitational perturbations.
Now, we have determined $D_{\Phi}$ in closed form at four dimensions \cite{gui} 
\begin{equation}
\label{10}
D_{\Phi}=\frac{{\rm e}^{i(\varphi -\varphi_{0})\gamma}\sinh [(\eta (1-\gamma )/B]+
{\rm e}^{-i(\varphi -\varphi_{0})(1-\gamma )}\sinh (\eta \gamma /B)}
{8\pi^{2}B\rho \rho_{0}\sinh \eta \, [\cosh (\eta /B)-\cos (\varphi -\varphi_{0}) ]} 
\quad {\rm with} \; \gamma =[\Phi ]
\end{equation}
\begin{equation}
\label{10a}
{\rm where} \quad 
\cosh \eta =
\frac{\rho^{2}+\rho_{0}^{2}+(x^{1}-x_{0}^{1})^{2}+(x^{2}-x_{0}^{2})^{2}}
{2\rho \rho_{0}}
\end{equation}
therefore we obtain the general expression of $S_{E}$ and
$G_{E\mu \mu_{0}}$ in terms of $B$ by formulas (\ref{4}) and (\ref{5}).
 For a cosmic string the physical value of $B$
is such that $B\leq 1$. In this case $[1/2-B/2]=1/2-B/2$ and $[B]=B$ and then 
 $S_{E}$ reduces to the result of Frolov
and Serebriany \cite{fro} and $G_{E\mu \mu_{0}}$ to the one of Allen {\em et al}
\cite{all}. 

Another physical case is the one of the field theories at finite
temperature $T$ on the Rindler space which is described by the Euclidean metric
\begin{equation}
\label{12}
ds^{2}=d\xi^{2}+\xi^{2}d\tau^{2}+(dx^{1})^{2}+(dx^{2})^{2}
\end{equation}
in a coordinate system $(\xi ,\tau ,x^{1},x^{2})$ such that $\xi >0$ and
$0\leq \tau <2\pi$. To analyse these theories, it is necessary to know the spin-weighted
Green's functions antiperiodic in $\tau$ for $s$ half-integer and periodic in 
$\tau$ for $s$ integer, the period being $\beta =1/kT$. The equivalence with
the previous problem is obtained by setting
\begin{equation}
\label{13}
\rho =\xi \quad {\rm and} \quad
\varphi =\frac{2\pi}{\beta}\tau \quad {\rm with} \quad 0\leq \tau <\beta 
\quad {\rm and} \quad B=\frac{\beta}{2\pi}
\end{equation}
In notations (\ref{13}), we rewrite the Green's function $D_{\Phi}$
\begin{equation}
\label{20}
D_{\Phi}=\frac{{\rm e}^{2i\pi \gamma (\tau -\tau_{0})/\beta}
\sinh [2\pi \eta (1-\gamma )/\beta ]+
{\rm e}^{-2i\pi (1-\gamma )(\tau -\tau_{0})/\beta}\sinh (2\pi \eta \gamma /\beta )}
{4\pi \beta \xi \xi_{0} \sinh \eta \, [\cosh (2\pi \eta /\beta )- 
\cos (2\pi (\tau -\tau_{0})/\beta )]}  
\quad {\rm with} \;  \gamma =[\Phi]
\end{equation}
and also formulas (\ref{9})
\begin{eqnarray}
\label{21}
\nonumber & & {\cal G}_{(s)}=\exp [-is(\tau -\tau_{0})]D_{1/2+\beta s/2\pi}
(\xi ,\tau -\tau_{0},x^{1},x^{2}) \quad s=\pm 1/2, \pm 3/2, \dots \\
& & {\cal G}_{(s)}=\exp [-is(\tau -\tau_{0})]D_{\beta s/2\pi}
(\xi ,\tau -\tau_{0},x^{1},x^{2}) \quad s=0,\pm 1,\dots
\end{eqnarray}
For the massless spinor Green's function at finite temperature, we obtain 
\begin{eqnarray}
\label{14}
\nonumber & & S_{\beta}=\left( e_{\underline{a}}^{\mu}\gamma^{\underline{a}}
\partial_{\mu} +\frac{\gamma^{\underline{1}}}{2\xi} \right) \left[
I \Re \left( \exp [i(\tau -\tau_{0})/2] \, D_{1/2-\beta /4\pi}
(\xi ,\tau -\tau_{0},x^{1},x^{2})  \right) \right. \\
& & \left. +\gamma^{\underline{1}}\gamma^{\underline{2}} \Im \left( 
\exp [i(\tau -\tau_{0})/2] \, D_{1/2-\beta /4\pi}(\xi ,\tau -\tau_{0},x^{1},x^{2})
\right) \right]
\end{eqnarray}
After some calculations, we see that  (\ref{14}) gives
for $\beta \leq 2\pi$ the known result \cite{van}.
For the electromagnetic field, we obtain the components of the vector Green's function
\begin{eqnarray}
\label{15}
\nonumber & & G_{\beta \xi \xi_{0}}=\Re \left( \exp [i(\tau -\tau_{0})] \,
\overline{D}_{\beta /2\pi}(\xi ,\tau -\tau_{0},x^{1},x^{2}) \right) \\
& & G_{\beta \tau \xi_{0}}=-\xi \Im \left( \exp [i(\tau -\tau_{0})] \,
\overline{D}_{\beta /2\pi}(\xi ,\tau -\tau_{0},x^{1},x^{2})  \right) \\
\nonumber & & G_{\beta\tau \tau_{0}}=\xi \xi_{0}G_{\beta \xi \xi_{0}} \quad {\rm and} 
\quad G_{\beta \xi \tau_{0}}=-\frac{\xi_{0}}{\xi}G_{\beta \tau \xi_{0}}
\end{eqnarray}
giving for $\beta \leq 2\pi$ the known result \cite{mor}.

In the Rindler space, the question of the validity of the expression 
of the thermal Green's functions  for $\beta >2\pi$
is crucial because for finding the thermal average of the energy-momentum 
tensor we have needed of the limit of the spin-weighted Green's functions at
zero temperature, {\em i.e.} $\beta \rightarrow \infty$. For the scalar field $s=0$,
we know the Euclidean Green's function ${\cal G}_{(0)\infty}$ at zero temperature
\cite{tro}
\begin{equation}
\label{23}
{\cal G}_{(0)\infty}=\frac{\eta}{4\pi^{2}\xi \xi_{0}\sinh \eta \, 
[\eta^{2}+(\tau -\tau_{0})^{2}]}
\end{equation}
With the aid of (\ref{23}), we can define a Green's function satisfying 
conditions (\ref{7}) by putting
\begin{equation}
\label{24}
\sum_{p=-\infty}^{+\infty}\exp (-2i\pi \Phi p){\cal G}_{(0)\infty}
(\xi ,\tau -\tau_{0}+\beta p,x^{1},x^{2})
\end{equation}
In virtue of the uniqueness of the solution,
we may identify (\ref{24}) with $D_{\Phi}$. Hence, we find
the spin-weighted Green's functions ${\cal G}_{(s)\infty}$ at zero temperature as
\begin{equation}
\label{25}
{\cal G}_{(s)\infty}=\exp [-is(\tau -\tau_{0})] \, {\cal G}_{(0)\infty}
(\xi ,\tau -\tau_{0},x^{1},x^{2})
\end{equation}
Taking into account (\ref{24}), we get the basic property for a Green's function
at zero temperature
\begin{eqnarray}
\label{26}
\nonumber & & {\cal G}_{(s)}=\sum_{p=-\infty}^{+\infty}(-1)^{p} \,
{\cal G}_{(s)\infty}(\xi ,\tau -\tau_{0}+\beta p,x^{1},x^{2}) 
\quad s=\pm 1/2, \pm 3/2,\dots \\
& & {\cal G}_{(s)}=\sum_{p=-\infty}^{+\infty}
{\cal G}_{(s)\infty}(\xi ,\tau -\tau_{0}+\beta p,x^{1},x^{2}) \quad s=0,\pm 1, \dots
\end{eqnarray}
We can then deduce the Green's functions at zero temperature for the massless
spin-$\frac{1}{2}$ field and for the electromagnetic field.

In the case of the scalar field, the canonical quantification at finite temperature
yields the thermal Feynman Green's function. A rotation of Wick in the Feynman
Green's function gives the Euclidean Green's function ${\cal G}_{(0)}$ which
is bounded in the limit where $\xi \rightarrow 0$.

For the massless spinor field and the electromagnetic field, several authors, 
working in the framework of the canonical quantification, obtain as expression  
of the Euclidean Green's functions for any $\beta$ the same form as the one
for $\beta \leq 2\pi$.
We emphasize that $D_{\Phi}$ given by (\ref{20}) vanishes at $\xi =0$. When we
replace $\gamma$ by $\Phi$, this expression gives also an Euclidean Green's function
but it blows up at $\xi =0$ for $\beta >2\pi$. So, it seems that the Euclidean 
Green's functions
derived from the thermal Feynman Green's functions are ill-defined at $\xi =0$
for $\beta >2\pi$.

{\bf Acknowlegdments}

I am grateful to Dr. Moretti for helpful discussions.

\end{document}